\documentclass[pre,aps,10pt,nosuperscriptaddress,twocolumn,floatfix]{revtex4-2}

\usepackage[nice]{nicefrac}
\usepackage{graphicx,color}
\usepackage{amssymb,bm,bbm}
\usepackage[tbtags]{amsmath}
\usepackage{braket}
\usepackage{bbold}
\usepackage{float}
\usepackage{physics}

\DeclareMathOperator*{\argmax}{argmax}

\usepackage[plainpages=false,pdfpagelabels,colorlinks=true,linkcolor=red,urlcolor=blue,citecolor=blue,pdftitle={},pdfauthor={},pdfdisplaydoctitle=true,pdfduplex=DuplexFlipLongEdge]{hyperref}

\definecolor{darkred}{rgb}{0.90,0.2,0.2}
\definecolor{darkgreen}{rgb}{0,0.60,.2}
\definecolor{darkblue}{rgb}{0.1,0.3,1}
\definecolor{grey}{cmyk}{0,0,0,0.25}
\definecolor{orange}{cmyk}{0,0.6,0.8,0}

\usepackage[T1]{fontenc}

\begin{document}
\title{Eigenstate entanglement entropy in Bose-Hubbard models}

\author{Gregor Medoš}
\author{Lev Vidmar}
\affiliation{Department of Theoretical Physics, J. Stefan Institute, SI-1000 Ljubljana, Slovenia}
\affiliation{Department of Physics, Faculty of Mathematics and Physics, University of Ljubljana, SI-1000 Ljubljana, Slovenia\looseness=-1}


\begin{abstract}
While the eigenstate entanglement entropy has been extensively studied for fermionic systems, much less is known about bosonic systems.
Here, we study the entanglement entropy of mid-spectrum eigenstates of Bose-Hubbard models, focusing on weakly disordered models with and without particle-number conservation, and contrasting them with the translationally-invariant model.
We analyze the volume-law and O(1) contributions to the entanglement entropy via the averages over mid-spectrum eigenstates and the corresponding distributions. 
We derive the volume-law coefficient of the entanglement entropy by generalizing the mean-field approach from [Phys.~Rev.~Lett. \textbf{119}, 220603 (2017)] to many-body systems with a tunable local bosonic cutoff, which agrees with previous analytical and numerical results from [Phys.~Rev.~B \textbf{110}, 235154 (2024)].
We show that the volume-law contribution to the entanglement entropy does not change upon breaking translational invariance via on-site disorder.
We then numerically study the role of the subleading O(1) contribution to the entanglement entropy.
We find that, in the particle-number conserving case, it exhibits a nontrivial dependence on the particle-number density and the local bosonic cutoff, while without particle-number conservation, results suggest the emergence of a universal O(1) contribution beyond the random pure state predictions.
\end{abstract}

\maketitle

\section{Introduction} \label{sec:introduction}

Entanglement entropy is a fundamental measure of entanglement in pure quantum states. 
Originating from quantum information theory, it has proven to be a powerful tool for characterizing quantum many-body states~\cite{srednicki_93, eisert_cramer_10, hastings_07}.
Considering a bipartition of a system into two subsystems, the scaling of the entanglement entropy with subsystem size -- either proportional to the boundary area or to the volume -- provides a simple distinction between ground states of condensed-matter systems and their highly excited eigenstates~\cite{eisert_cramer_10}.
Recently, it has been argued that a more detailed characterization of volume-law scaling can provide insight into a system’s ability to thermalize following a quantum quench~\cite{calabrese_cardy_05, calabrese_cardy_07, eisler_peschel_07, leblond_mallayya_19}.
This has shifted the focus toward studies of the entanglement entropies of highly excited eigenstates of quantum many-body Hamiltonians.

A reference point in these studies is the result by Page~\cite{page_93} for the average bipartite entanglement entropy $\expval{S_A}$ of two disjoint subsystems $A$ and $B$, with volumes $L_A$ and $L_B$, respectively, that together form a system of total volume $L=L_A+L_B$, prepared in a random pure state,
\begin{equation}\label{eq1}
    \expval{S_A} = \Psi(D_AD_B+1)-\Psi(D_B+1)- (D_A-1)/(2D_B)\;,
\end{equation}
where $\Psi$ is the logarithmic derivative of the $\Gamma$-function, and $D_A\leq D_B$ are the Hilbert space dimensions of subsystems $A$ and $B$, respectively.
Its asymptotic closed form is
\begin{equation} \label{def_SAinf}
    \expval{S_A} \asymp \ln D_A - D_A^2/(2D)\;.
\end{equation}
Hence, for a subsystem that consists of $L_A$ qudits with a local dimension $d_0$ and the Hilbert space dimension $D_A=d_0^{L_A}$, $\expval{S_A}$ consists of the leading volume-law term $\ln D_A = L_A \ln d_0$ that scales linearly with $L_A$, and the subleading term that either vanishes in the thermodynamic limit $L\to\infty$, or equals $-1/2$ if $D_A = \sqrt{D}$, which is true for $L_A=L/2$.

Recent studies of highly excited eigenstates of quantum many-body Hamiltonians have contributed to the following understanding of their entanglement entropies:
(i) Thermalizing systems exhibit the leading volume-law term identical to the one in Eq.~\eqref{def_SAinf}, i.e., their volume-law coefficient $\expval{S_A}/\ln D_A = 1$ is maximal~\cite{beugeling_andreanov_15, yang_chamon_15, vidmar_rigol_17, garrison_grover_18, dymarsky_lashkari_18, nakagawa_watanabe_18, huang_19, murthy_srednicki_19, miao_barthel_21, bianchi_hackl_22}.
This is in contrast to quadratic~\cite{vidmar_hackl_17, liu_chen_18, hackl_vidmar_19, lydzba_rigol_20, lydzba_rigol_21, bianchi_hackl_22, storms_singh_14} and integrable interacting systems~\cite{leblond_mallayya_19, swietek_kliczkowski_24}, in which the volume-law term may not be maximal and may depend on the subsystem fraction.
(ii) The subleading terms of interacting systems are much richer than the one predicted by Eq.~\eqref{def_SAinf}, and they carry information about the system's properties, such as symmetries~\cite{vidmar_rigol_17, bianchi_dona_19, bianchi_hackl_22, yauk_patil_24} and constraints~\cite{morampudi_chandran_20}.

An example that deviates from Eq.~\eqref{def_SAinf} is the case of particle-number conservation, which can be accounted for by considering random pure states in a fixed particle number sector~\cite{vidmar_rigol_17}.
It gives rise to the generalization of the Page formula to models with a global $U(1)$ symmetry, as is the case of particle-number conservation~\cite{bianchi_dona_19, bianchi_hackl_22},
\begin{equation}\label{eq2}
    \begin{split}
        \expval{S_A}_N &= \sum_{N_A} \frac{D_{A}^{(N_A)}D_B^{(N-N_A)}}{D^{(N)}}
        \bigg(\Psi(D^{(N)}+1) \\
        &- \Psi(\max\{D_A^{(N_A)} \,,\, D_B^{(N-N_A)}\}+1) \\
        &- \min\bigg\{\frac{D_A^{(N_A)}-1}{2D_B^{(N-N_A)}} \,,\, \frac{D_B^{(N-N_A)}-1}{2D_A^{(N_A)}}\bigg\}\bigg) \,,
    \end{split}
\end{equation}
where $D^{(N)}$ denotes the Hilbert space dimension of the full system with $N$ particles in volume $L$, $D_A^{(N_A)}$ is the Hilbert space dimension of subsystem $A$ with $N_A$ particles in volume $L_A$, and $D_B^{(N-N_A)}$ is the Hilbert space dimension of subsystem $B$, which contains $N_B=N-N_A$ particles in volume $L_B=L-L_A$.

In its asymptotic closed form, Eq.~\eqref{eq2} contains an O(1) contribution beyond the one predicted by Eq.~\eqref{def_SAinf},
\begin{eqnarray} \label{def_c1}
    c_1(f) = f/2+\ln(1-f)/2 \,,
\end{eqnarray}
where $f=L_A/L$ denotes the subsystem fraction.
This contribution to the entanglement entropy was first derived for interacting hard-core bosons or interacting two-levels systems in~\cite{vidmar_rigol_17}, and has recently been argued to generally apply to any physical system with $U(1)$ symmetry~\cite{yauk_patil_24}.
Moreover, different universal O(1) contributions were found for physical systems with $SU(2)$ symmetry~\cite{patil_hackl_23,bianchi_dona_24}.
Nevertheless, it is still debated whether other symmetries, such as the total energy conservation, also give rise to a universal O(1) contribution to the entanglement entropy~\cite{haque_mcclarty_22, kliczkowski_swietek_23, huang_24, rodriguez-nieva_jonay_24, langlett_rodriguez-nieva_25, langlett_jonay_25}.

The mid-spectrum eigenstate entanglement entropy of many-body quantum lattice systems has been extensively studied for systems with a two-dimensional local Hilbert space, such as spinless fermions, hard-core bosons, or equivalently, spin-\nicefrac{1}{2} systems~\cite{vidmar_rigol_17, bianchi_hackl_22, haque_mcclarty_22, kliczkowski_swietek_23, rodriguez-nieva_jonay_24, vidmar_hackl_17, lydzba_rigol_20, swietek_kliczkowski_24}.
However, much less focus has been devoted to bosonic systems, with a recent exception~\cite{yauk_patil_24}, where it was shown that the mid-spectrum eigenstate entanglement entropy of the translationally invariant Bose-Hubbard model~\cite{kollath_roux_10,cazalilla_citro_11} agrees with the $U(1)$ Page formula in Eq.~\eqref{eq2}.
Bosonic models such as the Bose-Hubbard models are important since they can be experimentally realized~\cite{lukin_rispoli_19, luschen_bordia_17, greiner_folling_08, rispoli_lukin_19, kondov_mcgehee_15, karamlou_rosen_24, choi_hild_16, roushan_neill_17, schreiber_hodgman_15, greiner_mandel_02transition, greiner_mandel_02revival, spielman_phillips_07, trotzky_chen_12} and even their entanglement entropy can be measured~\cite{sherson_weitenberg_10, kaufman_tai_16, islam_ma_15, daley_pichler_12}.
Theoretically, several studies have explored ergodicity in different versions of the Bose-Hubbard models, ranging from weak to strong disorder~\cite{sierant_zakrzewski_18,sierant_delande_17,hopjan_heidrich-meisner_20} and weak to strong interactions~\cite{pausch_carnio_21}.
For example, eigenstate entanglement entropy can be used as an indicator of ergodicity breakdown in Bose-Hubbard models at large disorder in which the crossover from volume-law to area-law scaling is expected, similar to the finite-size phenomenology of entanglement entropies in strongly-disordered spin-\nicefrac{1}{2} model Hamiltonians~\cite{sierant_lewenstein_25}.

In a recent work, the closed forms of Eqs.~\eqref{eq1} and~\eqref{eq2} have been generalized to any local Hilbert space dimension $d_0$~\cite{yauk_patil_24}.
This contributed to understanding the typical entanglement entropy for soft-core bosons, for which the number $n_j$ of bosons at site $j$ is upper bounded at each $j$ by a local bosonic cutoff $n_{\rm max}$, i.e., $n_j\leq n_{\rm max}$, such that $d_0=n_{\rm max}+1$.
True bosons correspond to the limit $n_{\rm max}\to\infty$, while the hard-core boson limit is set by $n_{\rm max}=1$.
The non-trivial volume-law coefficient of Eq.~\eqref{eq2} has been numerically tested for translationally-invariant systems for several $n_{\rm max}$ \cite{yauk_patil_24}.
Here, we explore the volume-law and O(1) contributions to the entanglement entropy in different versions of the Bose-Hubbard model, contrasting translationally invariant systems with weakly disordered ones, and cases with particle-number conservation with those where it is absent.
For the volume-law term, we introduce a pedagogical derivation based on the construction of random grandcanonical pure states, which generalizes the mean-field approach of Ref.~\cite{vidmar_rigol_17} to bosons and it yields results that agree with the recent result obtained in Ref.~\cite{yauk_patil_24}.
We then focus on numerical studies of the O(1) term beyond Eq.~\eqref{eq2}.
In particular, we find evidence that in the absence of particle-number conservation, there exists an additional O(1) term that may be universal, while in the particle-number conserving case, the O(1) term non-trivially depends on the particle number density $n=N/L$ and the local bosonic cutoff $n_{\rm max}$.

The paper is organized as follows.
In Sec.~\ref{sec2} we introduce the models and methods.
In Sec.~\ref{sec:volume} we derive the mean-field entanglement entropy for random grandcanonical pure states, which describes the leading volume-law term.
In Sec.~\ref{sec:TI} we examine the effect of breaking the translational invariance on the entanglement entropy.
In Sec.~\ref{sec:deviation} we observe a non-trivial dependence of the O(1) contributions to entanglement entropy in the case of particle-number conservation, and we contrast the results to the case without particle-number conservation.
We conclude in Sec.~\ref{sec:conclusions}.

\section{Models and Methods} \label{sec2}

We here provide some technical details of our study.
We introduce different versions of the Bose-Hubbard models in Sec.~\ref{sec:model}, we define the bipartite entanglement entropies for pure states with or without particle-number conservation in Sec.~\ref{sec:entropy}, and we discuss the numerical implementation in Sec.~\ref{sec:ed}.

\subsection{Bose-Hubbard Models} \label{sec:model}

\emph{Translationally-invariant Bose-Hubbard model.}
The translationally-invariant (TI) Bose-Hubbard model consists of two terms,
\begin{eqnarray} \label{def_H_TI}
    \hat H_{\rm TI}= \hat H_t + \hat H_U\;,
\end{eqnarray}
in which the first term describes the tight-binding nearest-neighbor hopping in the one-dimensional lattice with $L$ sites and periodic boundary conditions,
\begin{eqnarray}
    \hat H_t = -t \sum_{j=1}^L \left( \hat b_j^\dagger \hat b_{j+1} + \hat b_{j+1}^\dagger b_j \right)\;,
\end{eqnarray}
and the second term describes the on-site repulsion,
\begin{eqnarray}
    \hat H_U = \frac{U}{2} \sum_{j=1}^L \hat n_j \left( \hat n_j + 1 \right)\;,
\end{eqnarray}
where $\hat b_j^\dagger$ ($\hat b_j$) are bosonic creation (annihilation) operators at site $j$,
and $\hat n_j=\hat b_j^\dagger\hat b_j$ is the particle-number (site-occupation) operator.
We set the model parameters to $t=U=1$ such that the system described by Eq.~\eqref{def_H_TI} is in the quantum-chaotic regime~\cite{pausch_carnio_21}.

The model in Eq.~\eqref{def_H_TI} conserves the total number of particles due to an internal global $U(1)$ symmetry,
\begin{equation}
   \hat N = \sum_{j=1}^L \hat n_j\,,\;\;\;\
   \commutator{\hat H_{\rm TI}}{\hat N} = 0\,,
\end{equation}
Moreover, we must consider two additional lattice symmetries, i.e., translation and reflection symmetry.
The energy level statistics agree with the Gaussian Orthogonal Ensemble (GOE) predictions in each symmetry sector with given quantum numbers, the particle number $N$, the crystal momentum $K$, and the reflection parity $P_r$ for $K=0,\pi$.

\emph{Disordered Bose-Hubbard model.} 
Next, we introduce on-site disorder that breaks all lattice symmetries.
This is achieved via a random potential $\varepsilon_j \in [-1, 1]$ at each lattice site,
\begin{eqnarray}
    \hat H_W = \frac{W}{2}\sum_{j=1}^L \varepsilon_j \hat n_j\;,
\end{eqnarray}
where $W$ is the disorder strength parameter.
The disordered Bose-Hubbard model is then defined as
\begin{eqnarray} \label{def_H_DIS}
    \hat H_{\rm DIS}= \hat H_{\rm BH} + \hat H_W\,,
\end{eqnarray}
where $\hat H_{\rm BH}$ is given by $\hat H_{\rm TI}$ in Eq.~\eqref{def_H_TI}, adapted to the lattice with open boundary conditions.
Note that the model in Eq.~\eqref{def_H_DIS} still exhibits the particle-number conservation.
In this work we are interested in the weakly disordered regime.
In finite systems at large disorder, the level statistics ceases to comply with the GOE predictions~\cite{sierant_delande_17, sierant_zakrzewski_18, hopjan_heidrich-meisner_20}.

\emph{Generalized Bose-Hubbard model.} 
Finally, we introduce the generalized Bose-Hubbard model that breaks the particle-number conservation,
\begin{eqnarray} \label{def_H_GEN}
        \hat H_{\rm GEN}= \hat H_{\rm DIS} + \hat H_g \,.
\end{eqnarray}
This is achieved via the particle creation and annihilation terms that break the global $U(1)$ symmetry and therefore mix the particle-number sectors,
\begin{eqnarray} \label{def_H_g}
    \hat H_g = g \sum_{j=1}^L \left( \hat a_j + \hat a_j^\dagger \right)\,,
\end{eqnarray}
such that $\hat H_{\rm GEN}$ no longer commutes with $\hat N$.
To preserve Hermiticity, both processes in Eq.~\eqref{def_H_g} are equally weighted by the same parameter, and $g$ tunes the rate of particle production and destruction.
We set $g=1$ throughout our study.

Another important parameter that we set in the finite-size analyses of the models is the local bosonic cutoff $n_\mathrm{max}$, which restricts the number of bosons per site and hence truncates the local Hilbert space.
In the generalized Bose-Hubbard model from Eq.~\eqref{def_H_GEN} that mixes the particle-number sectors, one should in principle construct the full Hilbert space, which for true bosons is infinite. 
In our numerical calculations, we use $n_\mathrm{max}$ for every lattice site,
\begin{equation}
    n_j \leq n_{\rm max}\,,\;\; \quad \forall j\,.
\end{equation}
This corresponds to a spin chain with a finite spin quantum number, $S = n_\mathrm{max}/2$, and the local Hilbert space dimension $d_0=n_{\rm max}+1=2S+1$.
We refer to particles in this case as soft-core bosons. 
In the presence of particle-number conservation, truncation is not needed as long as $N\leq n_{\rm max}$. 
For soft-core bosons ($n_{\rm max}<\infty$), the largest particle-number sector corresponds to $n^* = n_{\rm max}/2$, but for true bosons ($n_{\rm max}\to\infty$), the largest particle number sector diverges.

\subsection{Bipartite entanglement entropy} \label{sec:entropy}

We define the von Neumann entropy of a subsystem as a measure for the bipartite entanglement entropy of a pure state $\ket{\psi}$.
From the reduced density operator $\hat \rho_A = \Tr_B \ketbra{\psi}{\psi}$ of a subsystem, denoted by $A$, we obtain the von Neumann entropy of that subsystem as
\begin{equation}\label{eq:vN}
    S_A(\psi) = -\Tr{\hat \rho_A \ln \hat \rho_A} \,,
\end{equation}
and identical entropy is obtained if the subsystem $A$ is replaced by the subsystem $B$.
When setting up the bipartition of the lattice into two sublattices of sizes $L_A$ and $L_B$, the Hilbert space is, in the absence of other symmetries such as the particle-number conservation, factorized via a tensor product,
\begin{equation} \label{def_tensor_product}
    \mathcal{H} = \mathcal{H}_A \otimes \mathcal{H}_B\,.
\end{equation}
Therefore, reshaping a state vector into an appropriately shaped state tensor is all one needs to perform a SVD decomposition.
From that one gains the square roots of the Schmidt coefficients $\lambda_j$, which are the eigenvalues of the reduced density matrix $\hat \rho_A$,
\begin{equation}
    S_A(\psi) = -\sum_{j=1}^{r_S}\lambda_j\ln\lambda_j\,,
\end{equation}
where $r_S$ is the number of non-zero $\lambda_j$, otherwise called the Schmidt rank.
The particle-number conservation, however, poses a constraint such that one cannot simply express a bipartition $L=L_A+L_B$ of the Hilbert space by a single tensor product as in Eq.~\eqref{def_tensor_product}.
One should rather express it as a direct sum of tensor products for each sector, corresponding to a unique way of dividing the number of particles $N=N_A+N_B$~\cite{bianchi_hackl_22},
\begin{equation}
    \hspace{-1cm}\mathcal H^{(N)} = \bigoplus_{N_A} \mathcal H_A^{(N_A)}\otimes\mathcal H_B^{(N-N_A)}\,.
\end{equation}
The structure of the Hilbert space hence makes the reduced density operator block diagonal, and the entanglement entropy,
\begin{equation}
    S_A(\psi) = \sum_{N_A=N_A^{\rm min}}^{N_A^{\rm max}} \Tr{\hat\rho_A^{(N_A)}\ln\hat\rho_A^{(N_A)}}\,,
\end{equation}
is hence a sum of contributions from sectors with a given number of particles $N_A$, which goes from $N_A^{\rm min}=\max\{0,\,(L-L_A)n_{\rm max}\}$ to $N_A^{\rm max}=\min\{N,\,L_An_{\rm max}\}$.

\subsection{Numerical implementation} \label{sec:ed}

\emph{Exact diagonalization.} 
We employ exact diagonalization (ED) to calculate exact Hamiltonian eigenstates.
For the translationally-invariant Hamiltonian, Eq.~\eqref{def_H_TI}, we obtain from $100$ up to $1000$ mid-spectrum eigenstates for each symmetry sector.
For the largest system sizes, we employ the polynomially filtered exact diagonalization (POLFED) algorithm~\cite{sierant_Lewenstain_polfed20} to calculate real symmetry sectors only, which have definite crystal momentum $0$ or $\pi$ and even or odd reflection parity. This gives four possible symmetry sectors for a system with even $L$, while for odd $L$ the $\pi$ crystal momentum sector does not exist, allowing for only the two symmetry sectors inside the $0$ crystal momentum sector.
For the disordered Hamiltonians, Eqs.~\eqref{def_H_DIS} and~\eqref{def_H_GEN}, the largest available system sizes are lower due to the absence of translational symmetry, and they depend on the local bosonic cutoff $n_{\rm max}$.
The largest system size for Hamiltonians with particle-number conservation, Eq.~\eqref{def_H_DIS}, is $L=14$ for $n_{\rm max}=2$, and $L=12$ for $n_{\rm max}>2$.
The largest system size for Hamiltonians without particle-number conservation, Eq.~\eqref{def_H_GEN}, is $L=12$ for $n_{\rm max}=2$, and $L=10$ for $n_{\rm max}>2$.
We consider $500$ realizations of the disordered Hamiltonians and for each Hamiltonian realization, we obtain from $500$ up to $1000$ eigenstates.

\emph{Statistics of eigenstate entanglement entropies.} 
Given a Hamiltonian eigenstate $\ket{\psi_n}$, we calculate the bipartite entanglement entropy $S_A^{(n)}=S_A(\psi_n)$ using Eq.~(\ref{eq:vN}) at the subsystem fraction $f=L_A/L=1/2$.
In the translationally-invariant (TI) case, $S_A^{(n,s)}$ denotes the entanglement entropy of an eigenstate labeled by $n$ in the symmetry sector labeled by $s$.
We calculate the mean entanglement entropy by averaging over $N_{\rm eig}$ mid-spectrum eigenstates within a given symmetry sector, followed by the average over all $N_s$ symmetry sectors,
\begin{equation} \label{def_S_TI}
    \overline{S_A}^{({\rm TI})} = \frac{1}{N_s} \sum_s S_A^{({\rm TI},s)} = \frac{1}{N_s}\frac{1}{N_{\rm eig}} \sum_s \sum_{n\in s} S_A^{(n,s)}\,,
\end{equation}
and the error is estimated by the standard deviation between the symmetry sectors,
\begin{equation} \label{def_S_TI_std}
    \delta \overline{S_A}^{({\rm TI})} = \sqrt{\mathrm{Var}_s(S_A^{({\rm TI},s)})}\,.
\end{equation}
Similarly, in the disordered case, $S_A^{(n,H)}$ denotes the entanglement entropy of an eigenstate labeled by $n$ for the Hamiltonian realization labeled by $H$.
We calculate the mean entanglement entropy by averaging over $N_{\rm eig}$ mid-spectrum eigenstates for a given Hamiltonian realization, followed by the average over $N_H$ Hamiltonian realizations,
\begin{equation}\label{AveS}
    \overline{S_A} = \frac{1}{N_H} \sum_{H} S_A^{(H)} = \frac{1}{N_H}\frac{1}{N_{\rm eig}} \sum_{H} \sum_{n\in H} S_A^{(n,H)}\,,
\end{equation}
and the error is given by the standard deviation between realizations,
\begin{equation}\label{ErrAveS}
    \delta \overline{S_A} = \sqrt{\mathrm{Var}_H(S_A^{(H)})}\,.
\end{equation}
We also calculate the distributions $P(S_A)$ of the numerically calculated eigenstate entanglement entropies.
We choose an appropriate bin width to obtain a smooth distribution, and we normalize the obtained distribution.
Using the kernel density estimation of the probability density function from the numerical distributions, we extract the statistical mode (mode for short),
\begin{equation}\label{MoS}
    \mathrm{Mo}(S_A) = \argmax_{S_A}\{P(S_A)\}\,,
\end{equation}
which represents the peak of the distribution.

\section{Mean-field calculation of entanglement entropies} \label{sec:volume}

The mean-field approximation for the bipartite entanglement entropy was introduced in~\cite{vidmar_rigol_17}.
It is based on the construction of random pure states with a targeted particle number $N$, or a targeted particle-number density $n$.
Once the reduced density matrix $\hat \rho_A$ of such a state is constructed, it is factorized in a particular way to separate the contributions from the average $\hat{\overline{\rho}}_A$ (that is a diagonal operator) and the corresponding fluctuations,
\begin{equation}
    \hat \rho_A = \hat{\overline{\rho}}_A(\hat I + \hat M)\,.
\end{equation}
Using such decomposition, the von Neumann entanglement entropy is
\begin{equation}
    S_A(\psi) = -\Tr\Big[
        \big(\hat{\overline{\rho}}_A + \hat{\overline{\rho}}_A \hat{M}\big) \big(\ln \hat{\overline{\rho}}_A + \ln(\hat{I} + \hat{M})\big)
    \Big]\,.
\end{equation}
With the term "mean-field" entanglement entropy we have in mind the approximation,
$S_A(\psi) \approx S_A^\mathrm{MF}$, in which the only contribution comes from the averaged reduced density matrix $\hat{\overline{\rho}}_A$,
\begin{equation}
    S_A^\mathrm{MF} = -\Tr{\hat{\overline{\rho}}_A \ln \hat{\overline{\rho}}_A}\,.
\end{equation}
In the case of hard-core bosons, it was shown that $S_A^\mathrm{MF}$ correctly describes the volume-law contribution to the entanglement entropy of mid-spectrum eigenstates with particle-number conservation~\cite{vidmar_rigol_17}.
Below, we generalize the mean-field approach by relaxing the hard-core constraint.

\emph{No particle-number conservation.}
In this case, the averaged reduced density matrix of random pure states will be a simple diagonal density matrix, $\hat{\overline{\rho}}_A = \hat{I}_A/D_A$,
where $D_A$ is the dimension of the Hilbert space for the subsystem $A$, given by $D_A = (n_\mathrm{max} + 1)^{L_A}$.
The corresponding mean-field entanglement entropy is then the volume-law contribution to the Page curve from Eq.~\eqref{def_SAinf},
\begin{equation}
    S_A^\mathrm{MF} = L_A \ln(n_\mathrm{max} + 1)\,.
\end{equation}
Hence, in the absence of particle-number conservation, the proportionality coefficient for the linear scaling with $L_A$ is the logarithm of the local Hilbert space dimension, $d_0 = n_{\rm max}+1$.
This is generally not the case if particle-number conservation is present~\cite{vidmar_rigol_17, bianchi_dona_19, bianchi_hackl_22}.

\emph{Particle-number conserving case.}
We now present the derivation of the volume-law term of the average entanglement entropy of random pure states with a fixed number of bosons.
To this end, we introduce a random grandcanonical pure state,
\begin{equation} \label{def_rgcps}
    \ket{\psi} = \sum_{a=1}^{D_A^{(N_A)}}\sum_{b=1}^{D_B^{(N-N_A)}} \frac{z_{ab}}{\sqrt{\mathcal{Z}(\mu)}} e^{\mu\hat{N}/2} \ket{a}\otimes\ket{b}\,,
\end{equation}
in which $z_{ab}$ are random coefficients that are independently sampled from the Gaussian distribution with zero mean and variance one, $\ket{a}$ and $\ket{b}$ correspond to the configurations of bosons in subsystems A and B, respectively, and the chemical potential $\mu$ is set such that the mean particle number is $N=Ln$. 
$\mathcal{Z}(\mu)$ is the grandcanonical partition function, which is equivalent to the generating function for the particle-number sector dimensions $D_N$~\cite{yauk_patil_24},
\begin{equation} \label{eqGeneratingFunction}
    \mathcal{Z}(\mu) = \zeta(\mu)^L\;,\;\;
    \mathcal \zeta(\mu) = 1 + e^\mu + \dots + e^{n_{\rm max}\mu}\,,
\end{equation}
where $\mu$ tunes the average particle number density,
\begin{equation}\label{eqLegendre}
    n = \dv{\ln\zeta(\mu)}{\mu}\,.
\end{equation}
Taking the partial trace of $\ketbra{\psi}{\psi}$, we obtain the reduced density matrix with a fixed average particle number,
\begin{equation}
    \hat{\rho}_A = \sum_{a,a'}^{D_A^{(N_A)}}\sum_{b=1}^{D_B^{(N-N_A)}}
    \frac{z_{ab}z_{a'b}^*}{\mathcal{Z}(\mu)} e^{\mu\hat{N}/2}\ketbra{a}{a'}e^{\mu\hat{N}/2}\,.
\end{equation}
The reduced density matrix can be decomposed into the averaged contribution,
\begin{equation} \label{def_barrhoA}
    \hat{\overline{\rho}}_A = \frac{e^{\mu\hat N_A}}{\zeta(\mu)^{L_A}}\,,
\end{equation}
and fluctuations.
Ignoring the fluctuations around the maximally mixed reduced density matrix, we retrieve the mean-field entanglement entropy,
\begin{equation} \label{def_SA_MF}
    S_A^{\rm MF} = F(n)L_A\,,
\end{equation}
where the proportionality coefficient for the linear scaling with $L_A$ is given by
\begin{equation}\label{eqVolumeLawCoefficient}
    F(n) = \ln\zeta(\mu) - \mu n\,.
\end{equation}
The expression in Eq.~\eqref{eqVolumeLawCoefficient} can be seen as the Legendre transform of the logarithm of the generating function \eqref{eqGeneratingFunction}.
Below we list examples of $F(n)$ for different values of $n_\mathrm{max}$:
\begin{enumerate}
    \item $n_\mathrm{max}=1$: hard-core bosons, fermions or spin-\nicefrac{1}{2},
    \begin{equation}
        F(n) = (n - 1)\ln(1 - n) - n\ln(n)\,.
    \end{equation}
    \item $n_\mathrm{max}=2$: soft-core bosons or spin-1,
    \begin{equation}\begin{split}
        F(n) &= (n - 2)\ln(2 - n) + (n - 1)\ln(2) \\
        &+ \ln(7 - 3n + \sqrt{1-3n(n-2)}) \\
        &- n\ln(n - 1 + \sqrt{1-3n(n-2)})\,.
    \end{split}\end{equation}
    \item $n_\mathrm{max}\to\infty$: true bosons or harmonic oscillators,
    \begin{equation}
        F(n) = (n + 1)\ln(1 + n) - n\ln(n)\,.
    \end{equation}
\end{enumerate}
We stress that strictly speaking, the random grandcanonical pure state in Eq.~\eqref{def_rgcps} does not contain a fixed particle number, but it only sets the mean particle-number density $n$.
Still, we argue that the resulting volume-law contribution to the entanglement entropy of random grandcanonical pure states is identical to those of the corresponding random {\it canonical} pure states; the same cannot be claimed for the subleading terms~\cite{bianchi_hackl_22}.
The advantage of this approach is that one does not need to deal with the canonical density matrix, which, in the case of random canonical pure states, should replace the grandcanonical density matrix $\hat{\overline{\rho}}_A$ in Eq.~\eqref{def_barrhoA}.

We note that the volume-law entanglement from the mean-field approach, see Eqs.~\eqref{def_SA_MF}-\eqref{eqVolumeLawCoefficient}, is identical to the result in Ref.~\cite{yauk_patil_24}.
There, using the saddle-point approximation for the particle-number sector dimensions $D^{(N)}$, $D_A^{(N_A)}$, $D_B^{(N_B)}$ that appear in the $U(1)$ Page formula in Eq.~\eqref{eq2},
\begin{equation}\label{eqScalingD}
    D^{(N)} \simeq e^{L F(n)}\,,
\end{equation}
allows one to calculate, in the limit $L\to\infty$, both the volume-law contribution as well as some subleading terms.
In particular, it gives rise to the closed-form expression for the entanglement entropy $\expval{S_A}_N$~\cite{yauk_patil_24}, which agrees with the previous result for two-dimensional local Hilbert spaces~\cite{bianchi_hackl_22},
\begin{equation}\label{eq2_closed_form}
        \begin{split}
        \expval{S_A}_N &\asymp F(n)L_A
        - \frac{\abs{F'(n)}}{\sqrt{2\pi\abs{F''(n)}}}\delta_{f,1/2}\sqrt{L} \\
        &+ \frac{1}{2}\left[f + \ln(1-f)\right]
        - \frac{1}{2}\delta_{f,1/2}\delta_{n,n^*}\,.
    \end{split}
\end{equation}
The maximum value of $F(n)$, and consequently of $D^{(N)}$, is at $n^*=n_{\rm max}/2$, for which $F'(n^*)=0$.
Performing a Taylor expansion around $n^*$ to second order in $n$, one gets a Gaussian approximation,
\begin{equation}\label{eq:GaussD}
    D^{(N)} \simeq e^{L F(n^*)} e^{-\frac{L}{2} \abs{F''(n^*)} (n-n^*)^2}\,.
\end{equation}
In the thermodynamic limit $L\to\infty$, one can use the Gaussian approximation of the Hilbert space dimension and notice that the width of the Gaussian distribution vanishes as a power law,
\begin{equation}
    \sigma = \frac{L^{-1/2}}{\sqrt{F''(n^*)}}\,,
\end{equation}
i.e., $\sigma\to0$ in the limit $L\to\infty$.
Moreover, the ratio between particle number sectors $\mathcal{H}^{(N)}$ will approach
\begin{equation}
    D^{(N)} / D^{(N^*)} \to 0\,,
\end{equation}
where $N^*=Ln^*=Ln_{\rm max}/2$.
In the absence of particle-number conservation, the entire Hilbert space is equally weighted.
Therefore, without particle-number conservation, the contribution from $\mathcal{H}^{(N)}$ with a particle number different from $N^*$ will be suppressed as $L\to\infty$.

In summary, we derived the volume-law coefficient, Eq.~\eqref{eqVolumeLawCoefficient}, for the particle-number conserving systems with arbitrary local Hilbert space dimension $d_0=n_{\rm max}+1$.
The mean-field approach provides a rather simple and physically motivated procedure for calculating the von Neumann entanglement entropy of the grandcanonical density matrix of a subsystem $A$.
It hence simplifies the analytical calculation by not needing to independently express the Hilbert space dimension of each particle number sector for subsystems $A$ and $B$.

The volume-law contribution to the entanglement entropy, as derived here, is expected to accurately describe the entanglement entropy of mid-spectrum eigenstates of bosonic Hamiltonians in the regime in which the system thermalizes.
This has been numerically shown in Ref.~\cite{yauk_patil_24} in systems with translation invariance.
In Secs.~\ref{sec:TI} and~\ref{sec:deviation} we go beyond the translationally-invariant models, and we numerically compare both the volume-law and O(1) contributions to the eigenstates entanglement entropies in different versions of the Bose-Hubbard models.

\section{Role of Translation invariance} \label{sec:TI}

We here compare the eigenstate entanglement entropies in the Bose-Hubbard models with and without translation invariance.
While in Sec.~\ref{sec:volume} we derived the analytical expression for the volume-law contribution to the entanglement entropy, we here test whether the same volume-law term applies to both translationally-invariant and disordered models.
This question is nontrivial since previous studies of entanglement entropies of many-body eigenstates of {\it noninteracting} Hamiltonians showed that translation invariance impacts the volume-law coefficient at nonvanishing subsystem fractions, $f>0$.
Specifically, the entanglement entropy of typical eigenstates of noninteracting disordered models such as the three-dimensional Anderson model~\cite{lydzba_rigol_21} complies with the predictions of Haar-random Gaussian states~\cite{lydzba_rigol_20, bianchi_hackl_21, bianchi_hackl_22}.
On the other hand, the volume-law coefficient of the entanglement entropy of typical eigenstates of translationally-invariant free fermions is lower~\cite{vidmar_hackl_17, bianchi_hackl_22, lisiecki_vidmar_25}.

Here we study whether similar effect emerges in the presence of interactions.
We focus on the subsystem fraction $f=1/2$ and compare the eigenstate entanglement entropies of the translationally-invariant model in Eq.~\eqref{def_H_TI} with those in the weakly disordered model in Eq.~\eqref{def_H_DIS}, setting $W=0.4$.

\begin{figure}[!t]
\centering
\includegraphics[width=\columnwidth]{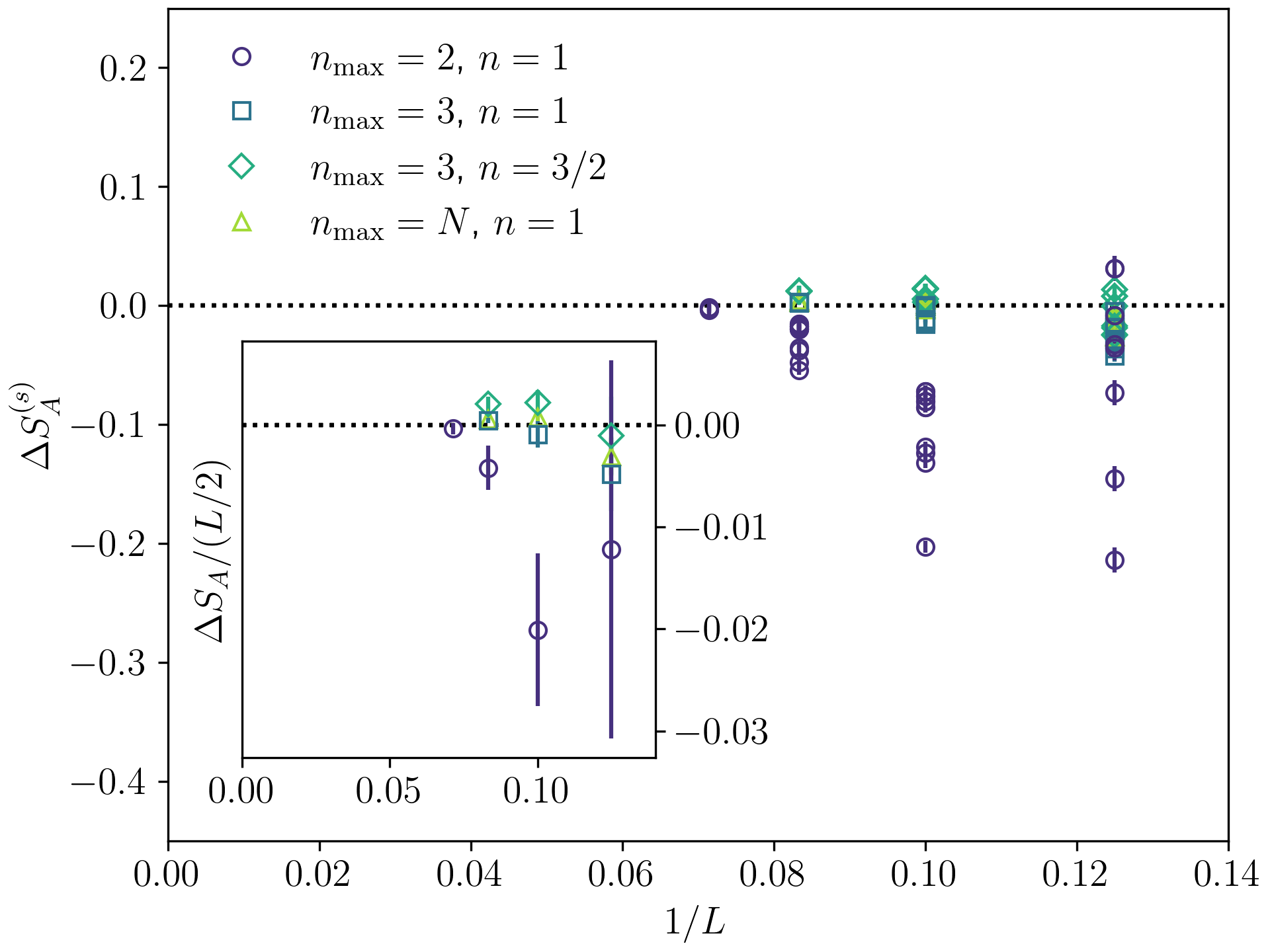}
\caption{
Differences of the mean eigenstate entanglement entropies between the translationally-invariant Bose-Hubbard model, see Eq.~\eqref{def_H_TI}, and the disordered Bose-Hubbard model, see Eq.~\eqref{def_H_DIS}, at $W=0.4$.
Main panel: symmetry resolved difference $\Delta S_A^{(s)}$, see Eq.~(\ref{DiffSymS}), for every symmetry sector $s$ of the translationally-invariant model, plotted vs $1/L$.
Symbols of different shapes correspond to results for different local bosonic cutoffs $n_{\rm max}$ and particle number densities $n$, while the results for different symmetry sectors at fixed $n_{\rm max}$ and $n$ are plotted with identical symbols.
The corresponding error bars (in most cases smaller than the symbol size) are given by the standard deviation of the average entanglement entropies between different disorder realizations, see $\delta \overline{S_A}$ from Eq.~\eqref{ErrAveS}.
Inset: scaled averages $\Delta S_A/(L/2)$, see Eq.~(\ref{DiffS}), where the average is taken over all symmetry sectors, and the error bars are defined in Eq.~(\ref{ErrDiffS}).
}
\label{fig1}
\end{figure}

We first study the deviations between the two models for each symmetry sector of the translationally-invariant model.
We define the symmetry-resolved deviation as
\begin{equation}\label{DiffSymS}
    \Delta S_A^{(s)} = S_A^{({\rm TI},s)} - \overline{S_A}\,,
\end{equation}
where $S_A^{({\rm TI},s)}$ and $\overline{S_A}$ are defined in Eqs.~\eqref{def_S_TI} and~\eqref{AveS}, respectively.
Results in the main panel of Fig.~\ref{fig1} show $ \Delta S_A^{(s)}$ vs $1/L$ at different local bosonic cutoffs $n_{\rm max}$ and particle number densities $n$.
Most of the values of deviations are very close to zero already for small system sizes $L$.
Symbols in Fig.~\ref{fig1} that exhibit the largest deviations correspond to the $k=0,\pi$ crystal momentum sectors, which were decomposed further into even and odd reflection parity sectors.
However, when $L$ increases, all the deviations become nearly indistinguishable from zero.

We also study the deviations between the averages in both models,
\begin{equation}\label{DiffS}
    \Delta S_A = \overline{S_A}^{({\rm TI})} - \overline{S_A}\,,
\end{equation}
where $\overline{S_A}^{({\rm TI})}$ is defined in Eq.~\eqref{def_S_TI}, and the standard deviation of Eq.~\eqref{DiffS} is calculated using Eqs.~\eqref{def_S_TI_std} and~\eqref{ErrAveS} as
\begin{equation}\label{ErrDiffS}
    \delta\Big(\Delta S_A\Big) = \sqrt{\Big(\delta\overline{S_A}^{({\rm TI})}\Big)^2 + \Big(\delta\overline{S_A}\Big)^2}\,.
\end{equation}
In the inset of Fig.~\ref{fig1} we plot the volume-law contribution to $\Delta S_A$, defined as $\Delta S_A/(L/2)$, versus $1/L$.
The obtained values of the volume-law contribution are very small already for small system sizes $L$, and they further decrease when $L$ increases.

Based on the observations in Fig.~\ref{fig1}, we therefore conclude that the translational symmetry in the Bose-Hubbard models does not modify the volume-law contribution to the average entanglement entropy.
Moreover, our results also suggest that the O(1) term is very similar (likely identical) when compared to the weakly disordered model.
We note that this behavior is different from the behavior in quadratic (free) fermion models, for which translation invariance yields deviations even on the level of the volume-law term~\cite{vidmar_hackl_17, hackl_vidmar_19, lydzba_rigol_20, bianchi_hackl_22}.

\section{O(1) contributions to entanglement entropies} \label{sec:deviation}

From now on, we only focus on the disordered Hamiltonians and study the accuracy of Eqs.~\eqref{eq1} and~\eqref{eq2} for the description of the average eigenstate entanglement entropies in the Bose-Hubbard models.
In particular, we ask whether there exist additional O(1) contributions to the eigenstate entanglement entropy that are not described by the random pure state predictions from Eqs.~\eqref{eq1} and~\eqref{eq2}.
This question has been recently addressed for interacting Hamiltonians of spinless fermions or spin-\nicefrac{1}{2} systems, and the results suggested that there may exist an O(1) term not captured by the random pure state predictions~\cite{haque_mcclarty_22, kliczkowski_swietek_23, huang_24, rodriguez-nieva_jonay_24}.
We separately consider cases with and without particle-number conservation in Secs.~\ref{sec:with_N} and~\ref{sec:no_N}, respectively.

We note that, as discussed in Introduction in Sec.~\ref{sec:introduction}, there exist O(1) contributions to the entanglement entropy even at the level of the random pure state prediction in Eqs.~\eqref{eq1} and~\eqref{eq2}.
For the case without particle-number conservation, see Eq.~\eqref{eq1}, the O(1) contribution is $-1/2$ at $f=1/2$, see Eq.~\eqref{def_SAinf}.
For the particle-number conserving case, see Eq.~\eqref{eq2}, the O(1) contribution contains the same term $-1/2$ at $f=1/2$ and $n=n^*=n_{\rm max}/2$, as well as an additional universal term due to particle-number conservation, see Eq.~\eqref{def_c1}. 
Here, when studying O(1) contributions to the eigenstate entanglement entropy of Hamiltonian systems, we are interested in the O(1) contributions beyond those contained in Eqs.~\eqref{eq1}--\eqref{def_c1}.

\subsection{Particle-number conserving case} \label{sec:with_N}

For the particle-number conserving case, we consider mid-spectrum eigenstates of the disordered Bose-Hubbard model Hamiltonian $\hat H_{\rm DIS}$ from Eq.~\eqref{def_H_DIS} at weak disorder $W=1$.

\begin{figure}[!t]
\centering
\includegraphics[width=\columnwidth]{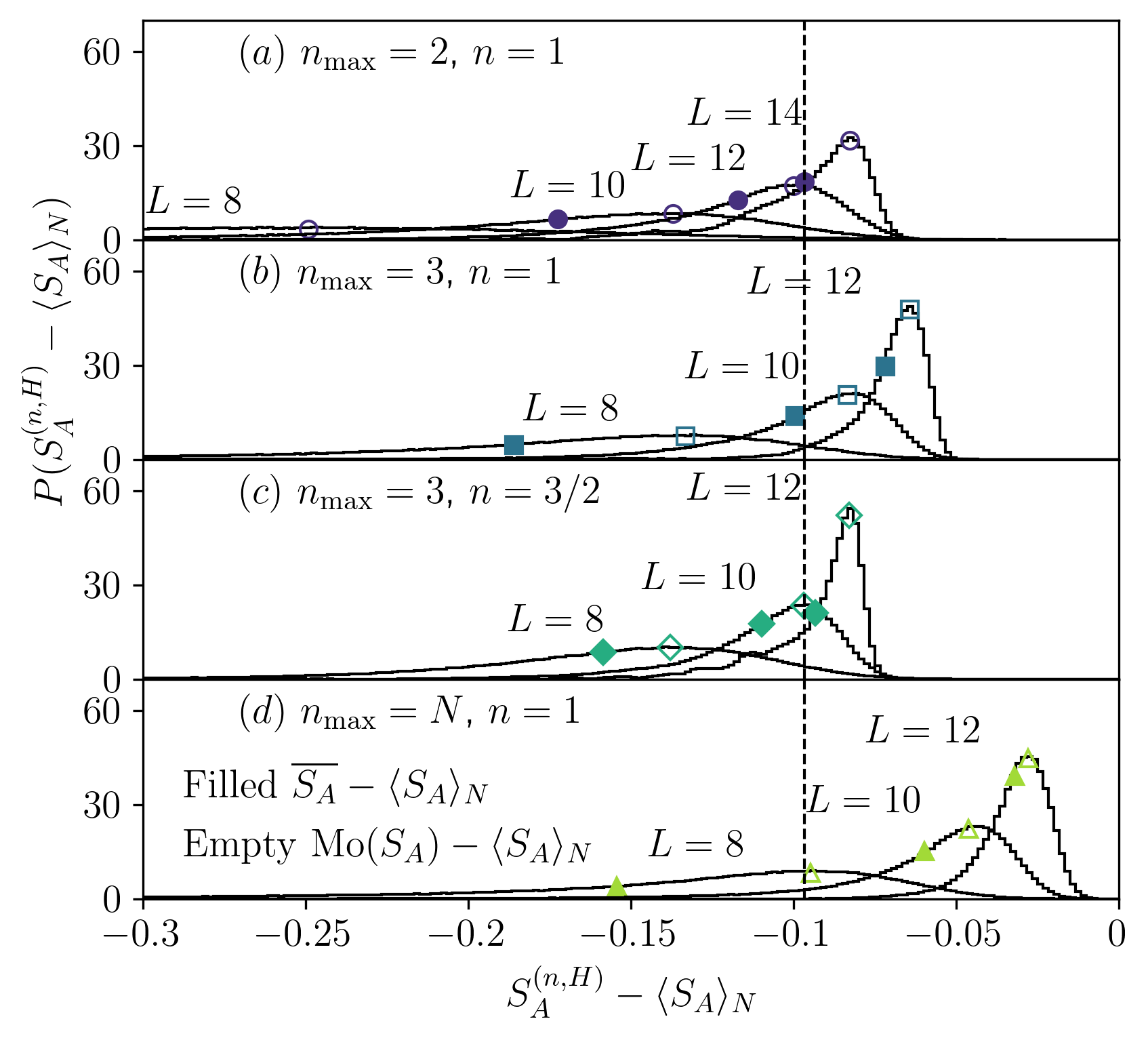}
\caption{
The distribution $P$ of the subtracted mid-spectrum eigenstate entanglement entropies $S_A^{(n,H)}-\expval{S_A}_N$, where $\expval{S_A}_N$ is the random pure state prediction from Eq.~\eqref{eq2}.
Symbols denote the subtracted means $\overline{S_A}-\expval{S_A}_N$, see Eq.~\eqref{AveS}, and the modes $\mathrm{Mo}(S_A)-\expval{S_A}_N$, see Eq.~\eqref{MoS}.
Results are shown for the particle-number conserving disordered Bose-Hubbard model in Eq.~\eqref{def_H_DIS}, at different values of the particle-number density $n$ and the local bosonic cutoff $n_{\rm max}$.
Vertical dashed lines denote the values $c_1(f=1/2)$, see Eq.~\eqref{def_c1}.
}
\label{fig2}
\end{figure}

In Fig.~\ref{fig2} we show the distributions $P$ of the mid-spectrum eigenstate entanglement entropies $S_A^{(n,H)}$, subtracted by the corresponding value $\expval{S_A}_N$ for the random pure states in Eq.~\eqref{eq2}. 
We show results for different local bosonic cutoffs $n_{\rm max}=2,3,N$ and different system sizes $L$, and we observe that in all cases, the subtracted entanglement entropies are negative.
The width of the distributions shrinks with increasing $L$, and both the mean and the mode exhibit a drift towards less negative values.

It is not immediately obvious whether the means and the modes of the distributions in Fig.~\ref{fig2} will shift to zero in the thermodynamic limit $L\to\infty$.
This tendency is, for the given system sizes, most clearly manifested for the case without the bosonic cutoff, i.e., for $n_{\rm max}=N$, see Fig.~\ref{fig2}(d).
At smaller $n_{\rm max}=2$ and $3$, see Figs.~\ref{fig2}(a)--\ref{fig2}(c), however, the situation is more subtle since the means and the modes depend on both $n_{\rm max}$ and the particle-number density $n$.
We observe that the deviations of the means and the modes from zero are largest when the particle-number density scales as $n=n^*=n_{\rm max}/2$, see Figs.~\ref{fig2}(a) and~\ref{fig2}(c).
Note that $n^*=n_{\rm max}/2$ corresponds to the dominant particle-number sector, see the discussion in Sec.~\ref{sec:volume}.
Therefore, it is likely that the further away the given particle-number sector is from the dominant sector, the faster the deviation from Eq.~\eqref{eq2} vanishes.

\begin{figure}[!t]
\centering
\includegraphics[width=\columnwidth]{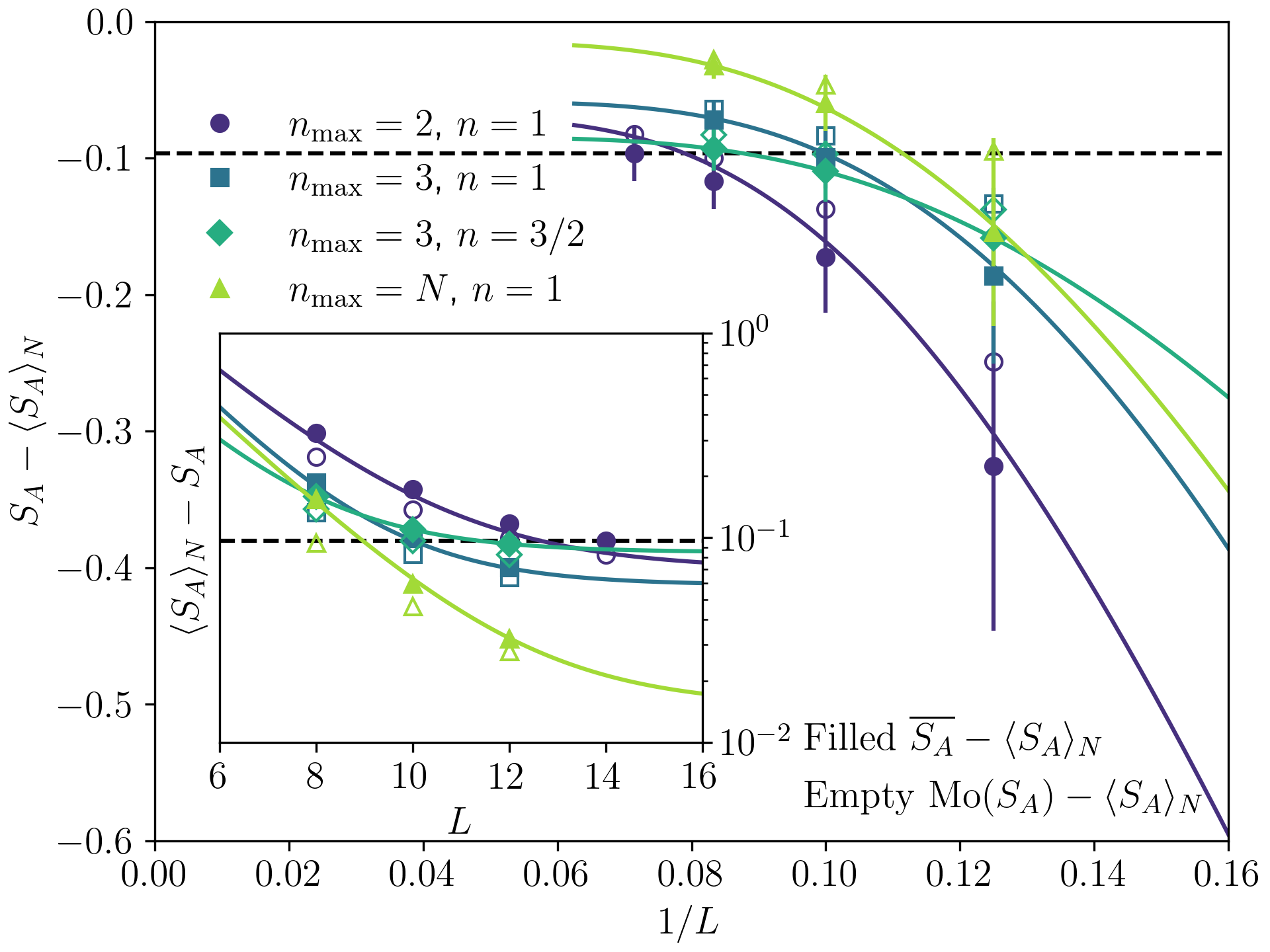}
\caption{
Subtracted means $\overline{S_A}-\expval{S_A}_N$, see Eq.~\eqref{AveS}, and the modes $\mathrm{Mo}(S_A)-\expval{S_A}_N$, see Eq.~\eqref{MoS}, of the mid-spectrum eigenstate entanglement entropies, shown versus $1/L$ in the main panel and versus $L$ in the inset.
$\expval{S_A}_N$ is the result for random pure states from Eq.~\eqref{eq2}.
Results are shown for the particle-number conserving disordered Bose-Hubbard model in Eq.~\eqref{def_H_DIS}, at different values of the particle-number density $n$ and the local bosonic cutoff $n_{\rm max}$.
Horizontal dashed lines denote the values $c_1(f=1/2)$, see Eq.~\eqref{def_c1}.
Solid lines are fits of the function $f(x)=A\exp(-Bx)+C$ to the numerical results for $\expval{S_A}_N-\overline{S_A}$.
We get $C=0.07, 0.06, 0.08, 0.02$ for $(n_{\rm max},n)=(2,1), (3,1), (3,\nicefrac{3}{2}), (N,1)$, respectively.
}
\label{fig3}
\end{figure}

Additional information about the finite-size scaling of the mid-spectrum entanglement entropies is given in Fig.~\ref{fig3}, in which we plot the mean~(\ref{AveS}) and the mode~(\ref{MoS}) of the distributions in Fig.~\ref{fig2}, versus $1/L$.
The fits, see the solid lines in Fig.~\ref{fig3}, may suggest that the values $\overline{S_A} - \expval{S_A}_N$ approach small nonzero values in the thermodynamic limit $L\to\infty$ that eventually depend on $n$ and $n_{\rm max}$.
On the other hand, since the deviations of $\overline{S_A} - \expval{S_A}_N$ from zero are very small, and the available system sizes are limited, one can also not exclude the possibility of approaching $\overline{S_A} \to \expval{S_A}_N$ as $L\to\infty$. 
It hence appears inconclusive to determine the asymptotic behavior when the thermodynamic limit is approached.

In case the O(1) differences between $\overline{S_A}$ and $\expval{S_A}_N$ persist in the thermodynamic limit $L\to\infty$, it is then relevant to ask whether there exist a possibly universal O(1) contribution not captured by the random pure state predictions in Eq.~\eqref{eq2}.
In Figs.~\ref{fig2} and~\ref{fig3}, see the dashed lines, we plot the constant $c_1(f=1/2)$ from Eq.~\eqref{def_c1}.
This constant was proposed to be relevant for the description of the additional O(1) contribution for fermionic and spin-\nicefrac{1}{2} systems~\cite{huang_24, rodriguez-nieva_jonay_24} and is conjectured to appear due to the total energy conservation.
We observe in Figs.~\ref{fig2} and~\ref{fig3} that in all particle-number conserving cases under consideration, both the means and the modes cross this value for the largest system sizes $L$.
However, when $n=n^*=n_{\rm max}/2$, see the circles and diamonds in Fig.~\ref{fig3}, the O(1) contribution is not considerably different from  $c_1(f=1/2)$ in Eq.~\eqref{def_c1}, and hence one cannot exclude that it ultimately converges to this value as $L\to\infty$.

In summary, we interpret the results of this section as follows.
While the O(1) contribution to the exact averages of the mid-spectrum entanglement entropies may or may not agree with the random pure state prediction from Eq.~\eqref{eq2}, their values at finite systems sizes depend nontrivially on $n_{\rm max}$ and $n$.
In case the deviations from the random pure state prediction persist in the thermodynamic limit $L\to\infty$, the interplay between $n_{\rm max}$ and $n$ makes the behavior in bosonic systems more subtle than the one observed in fermionic systems.

\subsection{Violation of particle-number conservation} \label{sec:no_N}

We now consider the generalized Bose-Hubbard model without particle-number conservation, $\hat H_{\rm GEN}$, defined in Eq.~\eqref{def_H_GEN} at weak disorder $W=1$.
In analogy to Sec.~\ref{sec:with_N} , we examine the distributions, means and modes of the mid-spectrum eigenstate entanglement entropies $S_A^{(n,H)}$, subtracted by the random pure state prediction $\expval{S_A}$ from Eq.~\eqref{eq1}.

\begin{figure}[!t]
\centering
\includegraphics[width=\columnwidth]{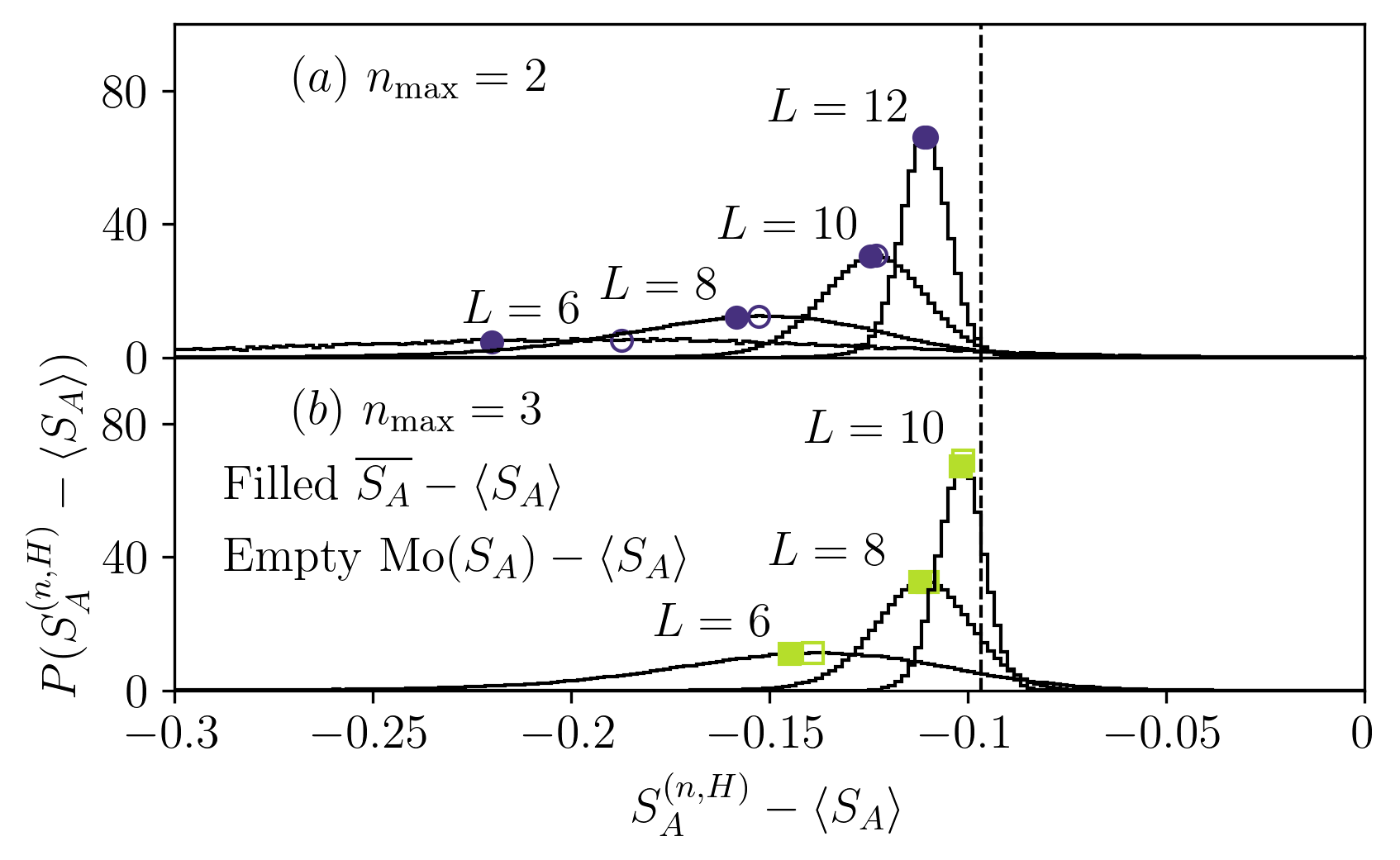}
\caption{
The distribution $P$ of the subtracted mid-spectrum eigenstate entanglement entropies $S_A^{(n,H)}-\expval{S_A}$, where $\expval{S_A}$ is the random pure state prediction from Eq.~\eqref{eq1}.
Symbols denote the subtracted means $\overline{S_A}-\expval{S_A}$, see Eq.~\eqref{AveS}, and the modes $\mathrm{Mo}(S_A)-\expval{S_A}$, see Eq.~\eqref{MoS}.
Results are shown for the generalized Bose-Hubbard model in Eq.~\eqref{def_H_GEN}, at different values of the local bosonic cutoff $n_{\rm max}$.
Vertical dashed lines denote the values $c_1(f=1/2)$, see Eq.~\eqref{def_c1}.
}
\label{fig4}
\end{figure}

The distributions $P$ of $S_A^{(n,H)} - \expval{S_A}$ are shown in Fig.~\ref{fig4}.
They share similarities with those for the particle-number conserving case shown in Fig.~\ref{fig2} in the sense that the width of the distributions narrows with increasing $L$, and the quantitative values of the mean and the mode depend, in the finite systems under investigation, on the local bosonic cutoff $n_{\rm max}$.
However, there is also an important difference compared to Fig.~\ref{fig2}: the mean and the mode remain lower than the value $c_1(f=1/2)$ from Eq.~\eqref{def_c1}, see the vertical dashed lines in Fig.~\ref{fig4}.

\begin{figure}[!t]
\centering
\includegraphics[width=\columnwidth]{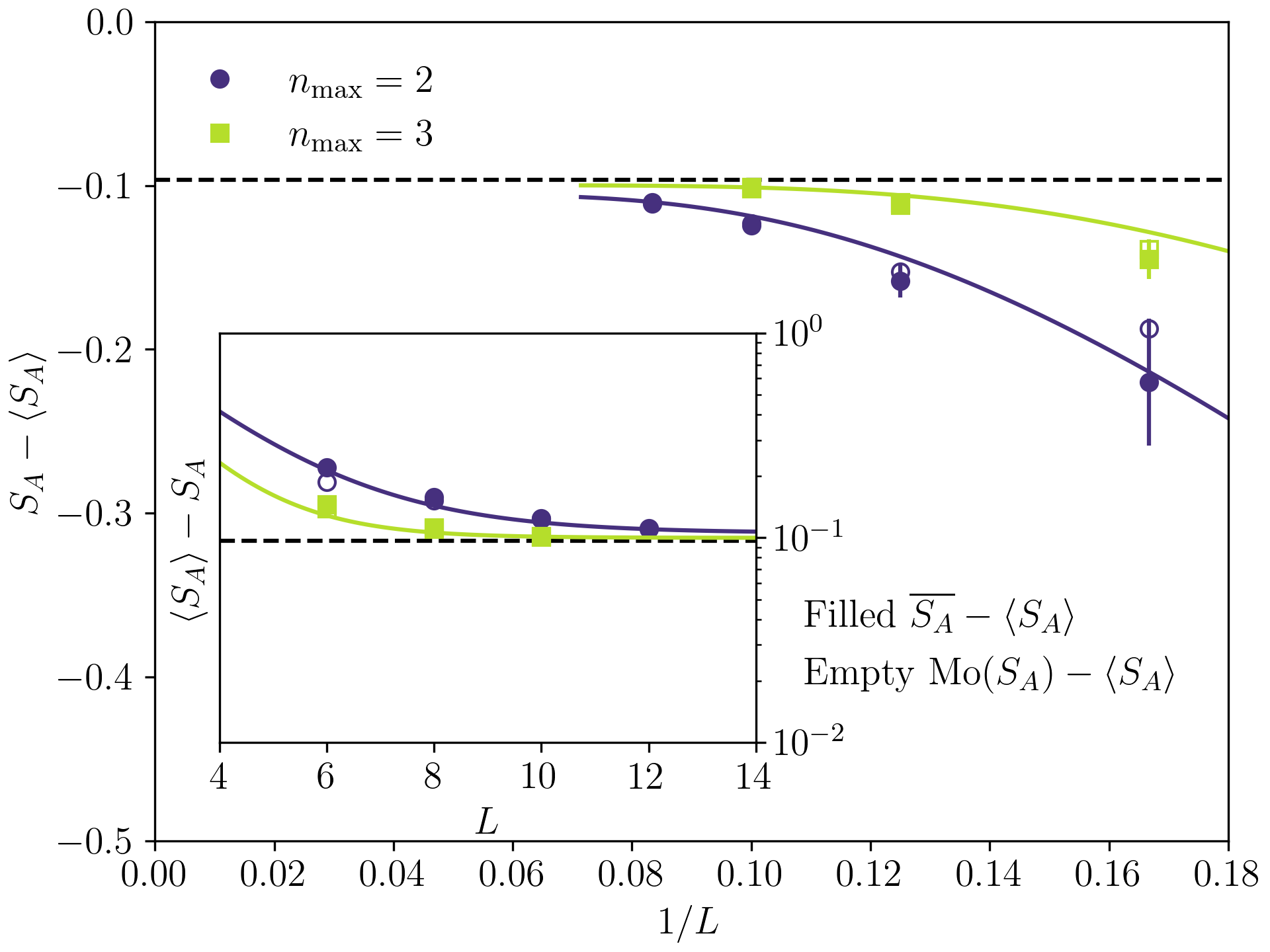}
\caption{
Subtracted means $\overline{S_A}-\expval{S_A}$, see Eq.~\eqref{AveS}, and the modes $\mathrm{Mo}(S_A)-\expval{S_A}$, see Eq.~\eqref{MoS}, of the mid-spectrum eigenstate entanglement entropies, shown versus $1/L$ in the main panel and versus $L$ in the inset.
$\expval{S_A}$ is the result for random pure states from Eq.~\eqref{eq1}.
Results are shown for the generalized Bose-Hubbard model in Eq.~\eqref{def_H_GEN}, at different values of the local bosonic cutoff $n_{\rm max}$.
Horizontal dashed lines denote the values $c_1(f=1/2)$, see Eq.~\eqref{def_c1}.
Solid lines are fits of the function $f(x)=A\exp(-Bx)+C$ to the numerical results for $\expval{S_A}-\overline{S_A}$.
We get $C=0.105, 0.100$ for $n_{\rm max}=2, 3$, respectively.
}
\label{fig5}
\end{figure}

In Fig.~\ref{fig5} we show the finite-size scaling of the subtracted means $\overline{S_A}-\expval{S_A}$ and the modes $\mathrm{Mo}(S_A)-\expval{S_A}$, versus $1/L$, which are marked by the symbols in the distributions in Fig.~\ref{fig4}.
The values exhibit a tendency to saturate to a non-zero value in the thermodynamic limit.
Moreover, for both choices of local bosonic cutoffs $n_{\rm max}=2$ and 3, they approach a similar value, which, to a good approximation, is given by $c_1(f=1/2)$ from Eq.~\eqref{def_c1}, see the horizontal dashed lines in Fig.~\ref{fig5}.

We hence interpret the results of this section as an indication that the O(1) difference between the exact numerical results $\overline{S_A}$ and the random pure state prediction $\expval{S_A}$ from Eq.~\eqref{eq1} persists in the thermodynamic limit.
Eventually, the difference may be given by $c_1(f=1/2)$ from Eq.~\eqref{def_c1}.
This O(1) contribution to the eigenstate entanglement entropies may share similarities with the one conjectured for spinless fermions and spin-\nicefrac{1}{2} systems~\cite{huang_24, rodriguez-nieva_jonay_24}.
However, we also note that in the one-dimensional spin-\nicefrac{1}{2} XYZ Hamiltonian, numerical results show that the absolute difference $|\overline{S_A} - \expval{S_A}|$ becomes smaller than $c_1(f=1/2)$ for systems that can be studied using exact diagonalization~\cite{kliczkowski_swietek_23}, i.e., they cross the horizontal dashed line in Fig.~\ref{fig5}.  
It hence appears that if such universal O(1) correction is indeed given by Eq.~\eqref{def_c1}, then the Bose-Hubbard Hamiltonians offer a clearer numerical playground for its detection.

\section{Conclusions} \label{sec:conclusions}

In this work, we studied the eigenstate entanglement entropies of Bose-Hubbard models with and without translational invariance, as well as with and without particle-number conservation.
We introduced the random grandcanonical pure states, which offer a rather simple evaluation of the volume-law contribution to the average entanglement entropy, thereby complementing previous work~\cite{yauk_patil_24}.
The resulting mean-field entanglement entropy can be seen as a generalization of the mean-field approach for hard-core bosons~\cite{vidmar_rigol_17} to bosons with an arbitrary local cutoff.

We found that the presence of translational invariance, when compared to the results of the weakly disordered model, does not modify the volume-law term, and likely also not the O(1) term of the entanglement entropy.
This behavior is in contrast to the behavior for non-interacting fermions, for which translational invariance modifies even the volume-law term~\cite{vidmar_hackl_17, lydzba_rigol_20, bianchi_hackl_22}.

Interesting behavior is found when comparing the particle-number conserving model with the one that violates particle-number conservation.
The differences arise on the level of a possible additional O(1) contribution to the entanglement entropy beyond the random pure state predictions.
We observed that the results in finite systems with particle-number conservation strongly depend on the choice of the particle-number density $n$ and the local bosonic cutoff $n_{\rm max}$.
In particular, the presence of $n_{\rm max}$ in bosonic systems gives rise to more subtle properties of entanglement entropies compared to two-level systems, e.g., spinless fermions or spins-\nicefrac{1}{2}.

In two-level systems, the dominant particle-number sector is $n^*=1/2$, while for bosons, it is $n^*=n_{\rm max}/2$.
Our results for the Bose-Hubbard models at $n=n^*$ are roughly consistent with the results for two-level systems, in the sense that there may exist an additional O(1) contribution to the entanglement entropy that is not described by the random pure state predictions.
Still, it is not entirely clear what is the value of this term.
In the case of no particle-number conservation, it is possible that the additional O(1) term is given by Eq.~\eqref{def_c1}, i.e., it is universal and identical to the one conjectured for two-level systems~\cite{huang_24, rodriguez-nieva_jonay_24, langlett_rodriguez-nieva_25}.
However, in the particle-number conserving case the modes of the entanglement entropy distributions in finite systems suggest that this O(1) term may be smaller than the value from Eq.~\eqref{def_c1}.
This result shares similarities with the observations for entanglement entropies in the spin-\nicefrac{1}{2} XYZ Hamiltonian~\cite{swietek_kliczkowski_24}, in which the O(1) term, for the largest system sizes under investigation, also became smaller than the value from Eq.~\eqref{def_c1}.

On the other hand, different behavior is observed when $n$ is different from $n^*$.
This case includes the paradigmatic particle-number conserving Bose-Hubbard model with $n=1$ and no local cutoff, i.e., $n_{\rm max}=N$.
In this case, it is likely that the additional O(1) term vanishes in the thermodynamic limit, i.e., the random pure state predictions for the mid-spectrum eigenstate entanglement entropies are likely exact even on the level of O(1) contributions.

\acknowledgements 
We acknowledge discussions with M. Rigol, and support from the Slovenian Research and Innovation Agency (ARIS), Research core funding Grants No.~P1-0044, N1-0273, J1-50005 and N1-0369, as well as the Consolidator Grant Boundary-101126364 of the European Research Council (ERC).
We gratefully acknowledge the High Performance Computing Research Infrastructure Eastern Region (HCP RIVR) consortium~\cite{vega1}
and European High Performance Computing Joint Undertaking (EuroHPC JU)~\cite{vega2}
for funding this research by providing computing resources of the HPC system Vega at the Institute of Information sciences~\cite{vega3}.



\bibliographystyle{biblev1}
\bibliography{acknowledgments,references}
\end{document}